\documentstyle[pre,aps]{revtex}

\tighten

\begin{document}


\draft
\title{
Energies and collapse times of 
symmetric and symmetry-breaking states 
of finite systems with a U(1) symmetry
}
\author{
Akira Shimizu\cite{email} and Takayuki Miyadera\cite{pa}
}
\address{
Department of Basic Science, University of Tokyo, 
Komaba, Meguro-ku, Tokyo 153-8902, Japan
}

\date{Received: 16 April 2001}

\maketitle
\begin{abstract}
We study quantum systems of volume $V$, 
which will exhibit the breaking of a U(1)
symmetry 
in the limit of $V \to \infty$, 
when $V$ is large but finite.
We estimate the energy difference between the 
`symmetric ground state' (SGS), 
which is the lowest-energy state that does not breaks the symmetry, 
and a `pure phase vacuum' (PPV), 
which approaches a symmetry-breaking vacuum as $V \to \infty$.
Under some natural postulates on the energy of the SGS, 
it is shown that 
PPVs always have a higher energy than the SGS,
and we derive a lower bound of the excess energy.
We argue that the lower bound 
is ${\cal O}(V^0)$, which becomes {\em much larger} than the
excitation energies of low-lying excited states for a large $V$.
We also discuss the collapse time of PPVs for interacting many bosons.
It is shown that the wave function collapses in a microscopic time
scale, because PPVs are not energy eigenstates.
We show, however, that for PPVs the expectation value of 
any observable, which is a finite polynomial 
of boson operators and their derivatives, 
does not collapse for a macroscopic time scale.
In this sense, the collapse time of PPVs is macroscopically
long.
\end{abstract}
\pacs{PACS numbers: 11.15.Ex, 05.30.-d, 05.10.-a, 03.75.Fi
}




The symmetry breaking (SB) is a key to understand
quantum systems of many degrees of freedom.
Although SB is {\it formally} defined for infinite systems, 
the physics of SBs in {\em finite} systems 
have been attracting much attention 
\cite{HL,miyashita,BLP,ADM,KT,SMprl2000,SMcluster,wright}
for the following reasons:
(i) Progress of experimental techniques enables 
one to observe and examine phase transitions in small systems, 
such as small magnets, small superconductors \cite{super}, 
liquid Helium in a small bubble \cite{He4bubble}, 
and laser-trapped atoms \cite{atom}, 
hence
SBs in finite systems should be studied seriously.
(ii) Although recent progress of computers enables one to 
obtain ground states of finite systems numerically, 
the relation between such ground states and 
SB ground states of {\it infinite} systems are non-trivial:
In a mean-field approximation, ground states of a finite system
of volume $V$
are degenerate (if it will exhibit a SB as $V \to \infty$),
and each ground state approaches 
a SB vacuum of the infinite system as $V \to \infty$.
We call such a state that
has a finite expectation value of an order parameter and 
approaches (i.e., well approximates) a SB vacuum of 
the infinite system as $V \to \infty$ 
`a pure phase vacuum' (PPV).
On the other hand, 
if one diagonalizes the Hamiltonian of the finite system exactly,
the energy spectrum is much different from that of 
a mean-field approximation.
The ground states are not necessarily degenerate.
Moreover, it often occurs that a symmetric state which does not break
the symmetry is a ground state, whereas PPVs have higher energies
\cite{HL,miyashita,BLP,ADM,KT,SMprl2000,SMcluster}, 
in the absence of a SB field, 
which is usually considered as an unphysical, 
artificial field for the breaking of a U(1) symmetry.
We call such a ground state 
the `symmetric ground state' (SGS).
In contrast to PPVs, 
the SGS does not approach a SB vacuum of 
the infinite system as $V \to \infty$.
Hence, for a SB to occur, 
the energy difference between PPVs and the SGS should be 
small enough.
Although the magnitude of this energy difference has been attracting 
much attention \cite{HL,miyashita,BLP,ADM,KT,SMprl2000,SMcluster}, 
definite conclusions have not yet been reached 
for the breaking of a U(1) symmetry.
For example, an exact calculation \cite{KT} gave 
only a rough estimate (see the discussion following Eq.\ (\ref{edCSIB})).

In this paper, we estimate much more strictly
the energy difference between PPVs and the SGS
for the breaking of a U(1) symmetry.
Under some natural postulates on the energy of the SGS,
we show that 
PPVs always have a higher energy than the SGS,
and that the excess energy is lower-bounded by 
$
\mu' \langle \delta N^2 \rangle /2V
$, where 
$
\mu'
$ is the derivative of the chemical potential with respect to 
the density $n \equiv \langle N \rangle / V$, 
and $\langle \delta N^2 \rangle$ denotes the fluctuation of $N$.
We further argue that this lower bound 
is ${\cal O}(V^0)$, which becomes {\em much larger} than the
excitation energies of low-lying excited states for a large $V$.
This should be contrasted with 
the breaking of the {\bf Z}$_2$ symmetry, 
for which the energy difference between PPVs and the SGS is
only ${\cal O}(V^{-1})$ \cite{HL}, 
which becomes {\em much smaller} than the excitation energies
in a three-dimensional space for a large $V$.
We also study the collapse time $t_{\rm coll}$ of PPVs
for the case of interacting many bosons.
It is shown that $t_{\rm coll} = {\cal O}(V^0)$ for 
the wave function, because PPVs are not energy eigenstates.
We show, however, that for PPVs $t_{\rm coll} = {\cal O}(\sqrt{V})$
for the expectation value of 
any observable, which is a finite polynomial 
of boson operators and their derivatives, 
if the degree of the polynomial is fixed independent of $V$.
In this sense, the collapse time of PPVs is macroscopically
long.

We consider a quantum system that has a U(1) symmetry, 
whose conserved charge $\hat N$ has integral eigenvalues
(in an appropriate unit).
We assume that the system is uniform, with the 
periodic boundary conditions, 
in order to eliminate additional complexities 
caused by non-uniform potentials or surface effects.
Since the system volume $V$ is finite, $\hat N$ is always well-defined, 
hence there exist simultaneous eigenstates 
$| N, \ell \rangle$ of $\hat H$ and $\hat N$;
\begin{eqnarray}
\hat H | N, \ell \rangle &=& E_{N,\ell} | N, \ell \rangle,
\\
\hat N | N, \ell \rangle &=& N | N, \ell \rangle,
\end{eqnarray}
where $\ell$ is a (set of) quantum number(s).
For each value of $N$, there exists the lowest-energy state 
$| N, {\rm G} \rangle$, which we assume is non-degenerate.
In general, $\hat N$ becomes the generator of the U(1) symmetry.
Hence, $| N, {\rm G} \rangle$ is the SGS.
We now make two basic postulates:
postulate 1 is the extensivity of the lowest eigenenergy;
\begin{equation}
E_{N, {\rm G}} = V \epsilon (N/V),
\label{p1}\end{equation}
where $\epsilon$ is a single-variable function of
the charge density $N/V$.
This postulate seems natural under our assumptions that 
the system is uniform and $| N, {\rm G} \rangle$ is non-degenerate,
as long as the charge does not induce a long-range force.
[Hence, care must be taken when the present results are applied 
to systems for which $N$ is the electric charge.]
By taking the limit $V \to \infty$, we can define $\epsilon (n)$ 
for every continuous value of $n$ ($=N/V$).
Using this $\epsilon (n)$ for a finite $V$ as well, we can regard $N$ in 
Eq.\ (\ref{p1}) as a continuous variable.
Hence, we can define
$ 
\mu(n) \equiv 
\epsilon' (n)
=
\frac{\partial}{\partial N}  E_{N, {\rm G}}.
$ 
Postulate 2 is
\begin{equation}
\mu'(n) \equiv 
\epsilon'' (n)
=
V \frac{\partial^2}{\partial N^2} E_{N, {\rm G}}
> 0,
\label{p2}\end{equation}
which also seems natural 
because thermodynamics requires that 
$\mu$ 
should be a non-decreasing function of $n$, for the system 
to be stable.
Although $\mu'=0$ for non-interacting particles
(such as a photon gas whose $\mu$ is always zero), we assume $\mu' > 0$ 
because we are not interested in such a trivial case.
For weakly-interacting many bosons with a repulsive interaction
(with the effective coupling constant $g>0$), for example,
these postulates are indeed satisfied;
$E_{N, {\rm G}} = V \epsilon (N/V)$, where
$ 
\epsilon (n) = g n^2/2 + \cdots,
$ 
hence
$ 
\mu' = g + \cdots >0.
$ 

When the system exhibits the breaking of the U(1) symmetry 
in the limit of $V \to \infty$ (while keeping the charge density 
finite), a SB state cannot be
$\lim_{V \to \infty} | N, {\rm G} \rangle$ or
$\lim_{V \to \infty} | N, \ell \rangle$, 
because they are eigenstates of $\hat N$ 
\cite{HL,miyashita,BLP,ADM,KT,SMprl2000,SMcluster}. 
Therefore, one should take superpositions of states 
with different charges in order to construct a PPV,
which approaches (i.e, well approximates) 
a SB vacuum as $V \to \infty$ 
\cite{HL,miyashita,BLP,ADM,KT,SMprl2000,SMcluster}.
If the quantum system (of a finite $V$) 
is perfectly closed, such superpositions are
forbidden for massive and/or charged particles
by the charge superselection rule, which requires that 
any pure state must be an eigenstate of $\hat N$ \cite{haag}.
However, if the quantum system is a subsystem of a larger system \cite{open}, 
we previously showed that one can associate a pure state, 
which is a coherent superposition of states with different charges,  
to the subsystem \cite{SMprl2000,SMssr}.
We investigate energy expectation values of such states, 
which in general have either of the following forms;
\begin{eqnarray}
| C \rangle
&\equiv&
\sum_{N}
C_N | N, {\rm G} \rangle,
\label{C}\\
| \tilde{C} \rangle
&\equiv&
\sum_{N, \ell}
\tilde{C}_{N, \ell} | N, \ell \rangle,
\label{tC}\end{eqnarray}
where $C_N$ and $\tilde{C}_{N, \ell}$ are coefficients, 
which are normalized as 
$\sum_N |C_N|^2 =1$ and 
$\sum_{N, \ell} |\tilde{C}_{N, \ell}|^2 = 1$, respectively.
We are interested in the case where the charge density 
$n$ ($= \langle N \rangle / V$)
approaches a finite value as $V \to \infty$.
(On the other hand, the case $n \to 0$ is rather trivial, while the case 
$n \to \infty$ is anomalous.)
Namely,  
\begin{equation}
\langle N \rangle = {\cal O}(V), 
\label{Norder}\end{equation}
where
$
\langle N \rangle
\equiv
\langle C | \hat N | C \rangle
$ 
or
$
\langle \tilde{C} | \hat N | \tilde{C} \rangle
$, and $n = \langle N \rangle/V$.
Furthermore, 
PPVs should be consistent with thermodynamics, 
according to which variances of extensive variables 
are of ${\cal O}(V)$ or smaller.
Hence, 
\begin{equation}
\langle \delta N^2 \rangle \leq {\cal O}(V),
\label{dN2order}\end{equation}
where 
$
\langle \delta N^2 \rangle
\equiv
\langle C | \delta \hat N^2 | C \rangle
$ 
or
$
\langle \tilde{C} | \delta \hat N^2 | \tilde{C} \rangle
$, where
$\delta \hat N \equiv \hat N - \langle N \rangle$.
Namely, $C_N$ and $\tilde{C}_{N, \ell}$ should 
localize in the $N$ space in such a way that 
Eqs.\ (\ref{Norder}) and (\ref{dN2order}) are satisfied.
Note that the phases of 
$C_N$ and $\tilde{C}_{N, \ell}$ are irrelevant to 
$\langle N \rangle$, 
$\langle \delta N^2 \rangle$, 
and 
the energy expectation values,
$
\langle C | \hat H | C \rangle
=
\sum_N |C_N|^2 E_{N,{\rm G}}
$ and $
\langle \tilde{C} | \hat H | \tilde{C} \rangle
=
\sum_{N, \ell} |C_N|^2 E_{N, \ell}
$.

For $N = \langle N \rangle + \Delta N$, 
where $(\Delta N)^2 \leq {\cal O}(V)$, 
Eq.\ (\ref{p1}) can be expanded as
\begin{equation}
E_{N,{\rm G}}
=
V \left[
\epsilon\left(\frac{\langle N \rangle}{V}\right)
+\frac{\Delta N}{V} \mu \left(\frac{\langle N \rangle}{V}\right)
+\frac{1}{2}\left(\frac{\Delta N}{V}\right)^2
\mu' \left(\frac{\langle N \rangle}{V}\right)
+ 
{\cal O} \left( \frac{1}{V^{3/2}} \right)
\right].
\label{expansion}\end{equation}
We neglect the higher-order term, 
$
V {\cal O} (1/V^{3/2})
$, 
in the following analysis.
Then, we can easily show that
\begin{equation}
\langle C | \hat H | C \rangle
=
E_{\langle N \rangle, {\rm G}}
+
\frac{\langle \delta N^2 \rangle}{2V} 
\mu'\left(\frac{\langle N \rangle}{V}\right).
\label{CHC}\end{equation}
Since the last term is positive because of postulate 2, 
we conclude that {\em 
$| C \rangle$ always has a higher energy 
than $| N, {\rm G} \rangle$ if they have the same 
value of $\langle N \rangle$
}.
Here, it is crucial to fix the value of $\langle N \rangle$
for the comparison. [Otherwise, either state could have a higher energy 
because of the linear term in Eq.\ (\ref{expansion}).]
Note that formula (\ref{CHC}), although very simple, gives 
the energy of a general state of the form (\ref{C})
very precisely, with the error being only 
${\cal O} (1/V^{1/2})$.
Note also that 
the energy expectation value is determined 
only by $\langle N \rangle$ and $\langle \delta N^2 \rangle$
if the functional forms of $E_{N, {\rm G}}$ and $\mu'(n)$ are given.

For a more general state (\ref{tC}), we derive an inequality.
Let $C'_{N}$'s be some coefficients which satisfy
$
|C'_{N}|^2 = \sum_{\ell} |\tilde{C}_{N, \ell}|^2
$.
For 
$| C' \rangle \equiv \sum_N C'_N | N, {\rm G} \rangle$, 
\begin{equation}
\langle \tilde{C} | \hat H | \tilde{C} \rangle
-
\langle C' | \hat H | C' \rangle
=
\sum_{N, \ell} |\tilde{C}_{N, \ell}|^2 (E_{N, \ell} - E_{N, {\rm G}})
\geq
0,
\end{equation}
where the equality holds iff 
$\tilde{C}_{N, \ell} = 0$ for all $\ell \neq {\rm G}$.
Applying Eq.\ (\ref{CHC}) to $\langle C' | \hat H | C' \rangle$,
and noting that 
$
\langle \tilde{C} | \hat N | \tilde{C} \rangle
=
\langle C' | \hat N | C' \rangle
$ 
and
$
\langle \tilde{C} | \delta \hat N^2 | \tilde{C} \rangle
=
\langle C' | \delta \hat N^2 | C' \rangle
$,
we obtain
\begin{equation}
\langle \tilde{C} | \hat H | \tilde{C} \rangle
\geq
E_{\langle N \rangle, {\rm G}}
+
\frac{\langle \delta N^2 \rangle}{2V} 
\mu'\left(\frac{\langle N \rangle}{V}\right).
\label{tCHtC}\end{equation}
When
$| \tilde{C} \rangle$ and $| C \rangle$ 
have the same values of 
$\langle N \rangle$ and $\langle \delta N^2 \rangle$, 
Eqs.\ (\ref{CHC}) and (\ref{tCHtC}) can be combined as the simple formula;
\begin{equation}
\langle \tilde{C} | \hat H | \tilde{C} \rangle
\geq
\langle C | \hat H | C \rangle
=
E_{\langle N \rangle, {\rm G}}
+
\frac{\langle \delta N^2 \rangle}{2V} 
\mu'\left(\frac{\langle N \rangle}{V}\right).
\label{combine}\end{equation}
It is easy to show the similar result for 
$
\hat K \equiv \hat H - \mu \hat N
$ (where $\mu$ here is a constant);
\begin{equation}
\langle \tilde{C} | \hat K | \tilde{C} \rangle
\geq
\langle C | \hat K | C \rangle
=
K_{\langle N \rangle, {\rm G}}
+
\frac{\langle \delta N^2 \rangle}{2V} 
\epsilon'' \left(\frac{\langle N \rangle}{V}\right),
\label{Kcombine}\end{equation}
where $K_{N, {\rm G}} \equiv E_{N, {\rm G}} - \mu N$. 
Hence, the following results are applicable to the expectation
values of $\hat K$ as well, if we replace 
$K_{N, {\rm G}}$ and $\mu'$ with $E_{N, {\rm G}}$ and $\epsilon''$, 
respectively.

Since a PPV should take either form (\ref{C}) or (\ref{tC}),
we conclude that
(i) a PPV always has a higher energy than the SGS, 
and (ii) the excess energy is lower-bounded by 
$
\mu' \langle \delta N^2 \rangle /2V
$.
This is the first of the main results of this paper.
It should be mentioned that Ref.\ \cite{KT} tried to
give an {\em upper} bound of the energy 
{\em difference} between the SGS and
`low-lying states,' a linear combination of which is a PPV,
while our result gives a {\em lower} bound of
the energy {\em increase} of PPVs over the SGS.

We now estimate how the lower bound
$\mu' \langle \delta N^2 \rangle /2V$
behaves with increasing $V$.
For interacting many-bosons, 
we previously found the state vector of a PPV, which we called 
the coherent state of interacting bosons (CSIB) 
\cite{SMprl2000,SMcluster,cluster}.
This state vector, denoted by $|\alpha, {\rm G} \rangle$, 
has the form of Eq.\ (\ref{C}) with 
$ 
C_N = e^{-|\alpha|^2} \alpha^N/\sqrt{N!}.
$ 
Hence, 
$\langle \delta N^2 \rangle = \langle N \rangle$, 
and Eq.\ (\ref{CHC}) yields
\begin{equation}
\langle \alpha, {\rm G} | \hat H | \alpha, {\rm G} \rangle
-
E_{\langle N \rangle, {\rm G}}
=
(n/2) \mu'(n)
=
{\cal O}(V^0)
>0,
\label{edCSIB}\end{equation}
where the sign is determined by postulate 2 ($\mu' > 0$), and
$n = \langle N \rangle/V$.
On the other hand, if we apply
the inequality of Ref.\ \cite{KT} to the CSIB, we obtain
$
|
\langle \alpha, {\rm G} | \hat H | \alpha, {\rm G} \rangle
-
E_{\langle N \rangle, {\rm G}}
|
\leq
{\cal O}(\sqrt{V})
$, 
which diverges as $V \to \infty$.
Although there is no 
contradiction between the two results, the present result 
gives a much more accurate estimate:
the CSIB has a higher energy than the SGS 
by ${\cal O}(V^0)$, for the same value of $\langle N \rangle$.
Although one might expect that the energy increase would 
be a decreasing function of $V$ (as in the case of the breaking of the 
{\bf Z}$_2$ symmetry \cite{HL}), our result denies such a naive expectation.

For general systems which exhibit the 
breaking of a U(1) symmetry, 
we do not know the explicit forms of PPVs.
By virtue of relation (\ref{combine}), however, 
it is sufficient to estimate $\langle \delta N^2 \rangle$
for the estimation of the energy increase.
We argue that 
\begin{equation}
\langle \delta N^2 \rangle \sim \langle N \rangle
\label{dNsimN}\end{equation}
for general systems which exhibit the 
breaking of a U(1) symmetry, 
for the following reasons.
Macroscopic properties of 
PPVs must be stable against weak perturbations from 
environments.
The environments include those which exchange charges with the system.
We may apply the classical stochastic theory to estimate 
the stable distribution of the charges
because
(i) the phases of $C_N$ and $\tilde{C}_{N, \ell}$ are irrelevant to 
$\langle N \rangle$ and $\langle \delta N^2 \rangle$, 
and (ii) the phase coherence between the environments 
and the system may be negligible if the dephasing times of 
the environments are short enough.
Then, according to the classical stochastic theory, 
the steady-state distribution of the charges should satisfy
Eq.\ (\ref{dNsimN}) when charges are randomly exchanged with 
a huge environment.

From Eqs.\ (\ref{p2}), (\ref{Norder}), 
(\ref{combine}) and (\ref{dNsimN}), we obtain 
\begin{equation}
\langle {\rm PPV}| \hat H | {\rm PPV} \rangle
-
E_{\langle N \rangle, {\rm G}}
\geq
{\cal O}(V^0)
> 0,
\label{general}\end{equation}
for general systems which exhibit the 
breaking of a U(1) symmetry.
Note that this result is consistent with the theory of
SBs in {\em infinite} systems, 
according to which
PPVs have the same energy {\em density} as
the SGS \cite{rivers}.
In fact, Eq.\ (\ref{general}) yields
$
(\langle {\rm PPV}| \hat H | {\rm PPV} \rangle
-
E_{\langle N \rangle, {\rm G}})/V
\geq
{\cal O}(V^{-1})
\to 0$ as $V \to \infty$,
for the lower bound of the difference in the energy {\em densities}.
On the other hand, 
our result denies a naive expectation that 
the energies of PPVs and the SGS would be
`almost degenerate' in the sense that 
the energy difference would 
be a decreasing function of $V$.
Furthermore,  
the energy difference for a large $V$ becomes {\em much larger} than the
excitation energies of low-lying excited states, 
whose wavenumber $k \propto V^{-1/d}$ in a $d$-dimensional space, because
the excitation energy $\epsilon(k)$ behaves as 
$\epsilon(k) \propto |k| \propto V^{-1/d}$ for a linear dispersion, 
and $\epsilon(k) \propto |k|^2 = V^{-2/d}$ for a parabolic dispersion.
This should be contrasted with 
the breaking of the {\bf Z}$_2$ symmetry, 
for which the energy difference between PPVs and the SGS is 
only ${\cal O}(V^{-1})$ \cite{HL}, 
which becomes {\em much smaller} than the excitation energies
in a three-dimensional space for a large $V$.
This indicates, for example, that 
much more care is necessary for 
the breaking of the U(1) symmetry than for 
the breaking of the {\bf Z}$_2$ symmetry, 
when one tries to find a PPV by numerical calculations.

We finally discuss the collapse time of PPVs, by generalizing the 
discussion of Ref.\ \cite{wright}.
In general, PPVs are not an energy eigenstate, 
hence their wave functions deform 
in finite systems as time evolves.
Let $t_{\rm coll}$ be the {\em collapse time}, 
which is defined as the time scale at which 
this deformation becomes significant.
For example, 
for an initial ($t=0$) state of the form of Eq.\ (\ref{C}), it 
evolves as
$ 
| C \rangle \to
\sum_{N}
C_N e^{-i E_{N, {\rm G}} t} | N, {\rm G} \rangle
\equiv
| C; t \rangle,
$ 
where $\hbar$ is taken unity.
Since 
$E_{\langle N \rangle + \Delta N, {\rm G}}
-
E_{\langle N \rangle, {\rm G}}
=
\mu \Delta N 
+ \cdots
$ from Eq.\ (\ref{expansion}),
the difference of 
$| C; t \rangle$ from 
$| C \rangle$ becomes significant at 
$t = t_{\rm coll} \sim 1/\mu \sqrt{\langle \delta N^2 \rangle}$,
because the linear term $\mu \Delta N$ alters the relative phases
among $C_N$'s, 
except when $C_N$'s take some special forms.
As $V$ is increased, this time scale approaches zero if 
$\sqrt{\langle \delta N^2 \rangle}$
increases in proportion to $V$ as Eq.\ (\ref{dNsimN}).
On the other hand, PPVs must survive over a macroscopic time scale, 
i.e., $t_{\rm coll} \to \infty$ as $V \to \infty$ for PPVs.
To satisfy this condition,
$C_N$'s of PPVs must take some special forms.
For interacting many-bosons, for example, 
$C_N$'s of the CSIB indeed have special forms, 
for which the effect of the linear term $\mu \Delta N$
on $|\alpha, {\rm G}; t \rangle$
is completely absorbed as a time evolution of the single parameter $\alpha$.
In fact, 
\begin{eqnarray}
|\alpha, {\rm G}; t \rangle
&=& 
e^{-|\alpha|^2} \sum_N  \frac{\alpha^N}{\sqrt{N!}}
e^{-i [E_{\langle N \rangle, {\rm G}} + \mu \Delta N + 
\mu' (\Delta N)^2/2V]t}
| N, {\rm G} \rangle
\\
&=& 
e^{-i (E_{\langle N \rangle, {\rm G}} - \mu \langle N \rangle) t}
e^{-|\alpha|^2} \sum_N  \frac{(\alpha e^{-i \mu t})^N}{\sqrt{N!}}
e^{-i [\mu' (\Delta N)^2/2V]t}
| N, {\rm G} \rangle.
\end{eqnarray}
Since 
the prefactor 
$e^{-i (E_{\langle N \rangle, {\rm G}} - \mu \langle N \rangle) t}$
has no physical meaning,
we find that
$
|\alpha, {\rm G}; t \rangle
=
|\alpha e^{-i \mu t}, {\rm G} \rangle
$ if $\mu' = 0$.
Namely, the CSIB does not collapse at all if $\mu' = 0$; 
only the phase of $\alpha$ evolves with time as $\alpha e^{-i \mu t}$.
This result for $\mu'=0$ is well-known.
If $\mu' > 0$, on the other hand, 
the wave function $|\alpha, {\rm G} \rangle$ collapses 
at $t \sim V/\mu' (\Delta N)^2 \sim V/\mu' \langle \delta N^2 \rangle
= {\cal O}(V^0)$.
However, this does {\em not} necessarily mean that 
the expectation values of observables of interest alter 
in this time scale.
For example, if an observable is proportional to
the boson operator $\hat \psi$ or its derivative,
it detects the phase relation between adjacent coefficients,
$C_{N +1}$ and $C_N$.
The ratio of their phases evolves,
for $N = \langle N \rangle + \Delta N$, as 
\begin{equation}
\frac{
(e^{-i \mu t})^{N+1} e^{-i \mu' (\Delta N + 1)^2 t / 2V}
}{
(e^{-i \mu t})^{N} e^{-i \mu' (\Delta N)^2 t / 2V}
}
=
e^{-i \mu t} e^{-i \mu' t / 2V} e^{-i \mu' \Delta N t / V}.
\end{equation}
In the right-hand side, 
the first factor $e^{-i \mu t}$ can be absorbed as
the time evolution of the single parameter $\alpha \to \alpha e^{-i \mu t}$, 
whereas
the second factor $e^{-i \mu' t / 2V}$ is negligible 
because $\mu' / 2V = {\cal O}(1/V)$.
Hence, only the last factor $e^{-i \mu' \Delta N t / V}$ is relevant to 
the collapse time. We thus find
\begin{equation}
t_{\rm coll} 
\sim V/(\mu' \Delta N)
\sim V/(\mu' \sqrt{\langle \delta N^2 \rangle})
=
{\cal O}(\sqrt{V}).
\end{equation}
For general observables which are polynomials of degree $M$ 
of $\hat \psi$, $\hat \psi^\dagger$ and their derivatives, 
we obtain
\begin{equation}
t_{\rm coll} 
=
{\cal O}(\sqrt{V}/M).
\end{equation}
Hence, if the degree $M$ of the polynomial is fixed independent of $V$,
we again obtain $t_{\rm coll} = {\cal O}(\sqrt{V})$.
Since the expectation values of observables are 
relevant in quantum theory, we may conclude that
the collapse time of the CSIB is ${\cal O}(\sqrt{V})$, which is macroscopic 
in the sense that it diverges as $V \to \infty$,
although the collapse time of its wave function is ${\cal O}(V^0)$
\cite{similar}.
If, on the other hand, if $M$ is increased in proportion to $\sqrt{V}$,
then $t_{\rm coll} = {\cal O}(V^0)$.
Except for such an abnormal case (as $M \propto \sqrt{V}$), 
$t_{\rm coll}$ is macroscopic.
For more general systems with the breaking of U(1) symmetry,
we have not yet obtained definite conclusions on $t_{\rm coll}$, 
although we expect a situation similar to the case of interacting 
many bosons.
This may be a subject of future studies.


\begin{references}
\bibitem{open}
For example, the trapped BEC systems are not closed
because atoms escape from the trap via three-body collisons,
dipolar relaxations, and collions with atoms in background gases. Furthermore,
the trapped atoms consist of those in the ground state (BEC), a thermal cloud,
and (possibly) outer clouds; collisions among them occur continualy.
\bibitem{cluster}
The ground state with a defenite number of bosons, 
$| N, {\rm G} \rangle$, is the SGS that 
does not approach the correct vacuum of the infinite system as $V \to \infty$.
Namely, it was rigorously proved that a correct vacuum state must have the 
`cluster property' \cite{haag}, and it was shown for BEC systems that 
$| N, {\rm G} \rangle$ does not have this property, whereas the CSIB
has \cite{KT,SMcluster}.
This fact and the robustness of the CSIB against weak perturbations 
from environments \cite{SMprl2000} strongly suggest that for a sufficiently 
large $V$ the CSIB should be realized.
\bibitem{similar}
Similar results were previously obtained in Ref.\ \cite{wright}, 
which assumed almost-free bosons and $M=1$.
The present result is the generalization 
to general interacting many-boson systems and to general observables, 
for which $M \geq 1$.
\end{references}
\end{document}